# MODIFICATION OF LATTICE VIBRATIONS SPECTRUMS IN THE NANOPARTICLES OF PARADIBROMBENZOL/PARADIHLORBENZOL SOLID SOLUTION

**M.A.Korshunov**

*Institute of physics it. L. V.Kirenskogo of the Siberian separating of the Russian Academy of Sciences, 660036 Krasnoyarsk, Russia*
*e-mail: mkor@iph.krasn.ru*

Using Raman scattering, the solid solution nanoparticles of paradibrombenzol/paradihlorbenzol are investigated. It is found that in a spectrum of nanoparticles of solid solution frequencies of spectral lines are reduced in comparison with similar lines of a single crystal with the same concentration of components. Calculations of the lattice oscillations by Dyne's method are performed. It is shown that these modifications of frequencies can be caused by the increase of lattice parameters with the reduction of nanoparticles sizes.

Earlier it has been shown that frequencies of spectral lines of the lattice oscillations of a p-dibromobenzene in nanoparticles decrease. It is caused by magnification of parametres of a lattice [1]. Is of interest to explore, how the presence of impurities (paradihlorbenzol) influences the lattice oscillations of nanoparticles of a p-dibromobenzene. For this purpose, nanoparticles of solid solutions paradibrombenzol/paradihlorbenzol with a size of 200 nanometers with the given concentration of components have been prepared. The size of nanoparticles was determined by an electronic microscope. After this determination, the measurements of Raman spectrums were made on spectrometer Jobin Yvon T64000. Quantity of concentration of an impurity was determined by the relative intensity of the valence intramolecular vibrations. The spectral line in a p-dibromobenzene with frequency of 212.0 cm$^{-1}$ corresponds to a stretching vibration C-Br, and a line in paradihlorbenzol of 327.0 cm$^{-1}$ to stretching vibration C-Cl. In Fig.1 spectrums of intramolecular vibrations (in the range of 150 - 400 cm$^{-1}$) a p-dibromobenzene (1) paradihlorbenzol (2) single crystals of paradibrombenzol/paradihlorbenzol solid solution (3) and a studied nanoparticle of solid solution (4) are presented at the same concentration.

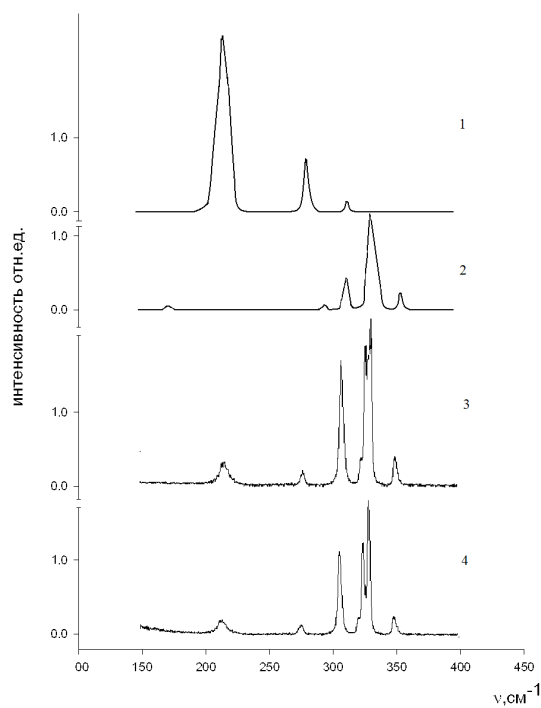

*Figure 1. Spectrums of intramolecular vibrations (in the range of 150 – 400 cm$^{-1}$).*

Concentration of molecules of the impurity, discovered in nanoparticles on the relative intensity of intramolecular vibrations, corresponds to 5% of a p-dibromobenzene. At the same time, the frequencies of lattice vibrations lines (ν) of volume solid solution and nanoparticles are different (see table).

| ν,cm$^{-1}$ nanoparticle | 25.4 | 44.8 | 47.3 | 51.7 | 92.13 | 100.0 |
|---|---|---|---|---|---|---|
| ν,cm$^{-1}$ single crystal | 26.84 | 46.15 | 48.0 | 53.5 | 92.3 | 100.3 |

Frequencies of nanoparticle are smaller. For a single crystal value of frequencies of this spectrum would correspond to a 13% concentration of a p-dibromobenzene. For an explanation of observable modifications of frequencies of lines calculations of spectrums of the lattice oscillations by Dyne's method [2] have been done. It is found that at 5% concentration of a p-dibromobenzene the spectrum of the lattice vibrations corresponds to a nanoparticle spectrum if lattice parameters are increased. The histogram of the calculated spectrum of the lattice vibrations of nanoparticles is presented in Fig.2(2) in comparison with the experimental spectrum of a nanoparticle in Fig.2(1) for the same concentration of components.

Thus, if to guess that size effects for studied sizes of nanoparticles in solid solutions affect intramolecular oscillations less, and the lattice oscillations that are more sensitive, apparently, there is a magnification of parametres of a lattice in studied nanoparticles.

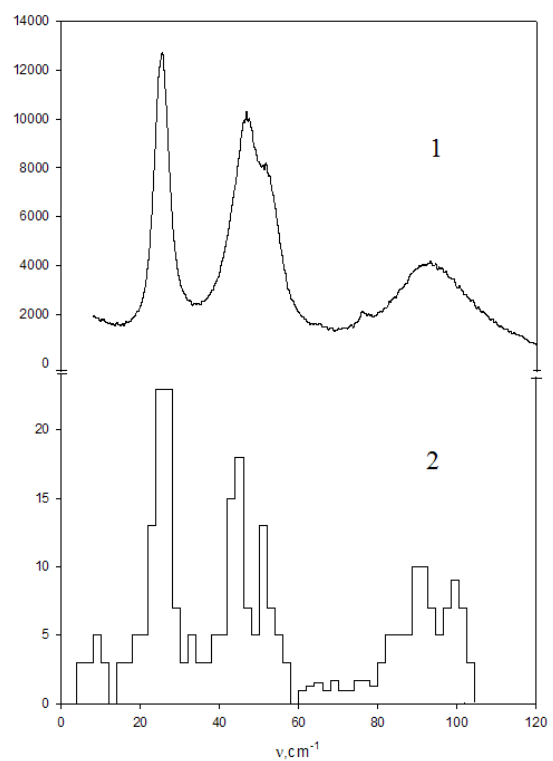

*Figure 2. The experimental spectrum (1) and the histogram of the calculated spectrum of the lattice oscillations of nanoparticles.*